\documentclass[prl,superscriptaddress,twocolumn,twoside,groupedaddress,floatfix,dvips]{revtex4}
\usepackage{amsfonts}
\usepackage{graphicx}
\usepackage[letterpaper]{hyperref}
\usepackage{times}
\usepackage{mathptmx}
\usepackage{amsmath}
\usepackage{amssymb}

\begin{document}
\preprint{}

\title{Multi-Dot Floating-Gates for Nonvolatile Semiconductor Memories \\-- Their Ion Beam Synthesis and Morphology}

\author{T. M\"{u}ller}
\email{T.Mueller@fz-rossendorf.de} \affiliation{Research Center Rossendorf, Institute of Ion Beam
Physics and Materials Research, PO Box 51 01 19, 01314 Dresden, Germany}
\author{C. Bonafos}
\affiliation{nMat Group, CNRS/CEMES, 29 Rue Jeanne Marvig, 31055 Toulouse, France}
\author{K.-H. Heinig}
\affiliation{Research Center Rossendorf, Institute of Ion Beam Physics and Materials Research, PO
Box 51 01 19, 01314 Dresden, Germany}
\author{M. Tenc\'{e}}
\affiliation{Laboratoire de Physique des Solides, Universit\'{e} Paris-Sud - UMR 8502, 91405 Orsay,
France}
\author{H. Coffin}
\affiliation{nMat Group, CNRS/CEMES, 29 Rue Jeanne Marvig, 31055 Toulouse, France}
\author{N. Cherkashin}
\altaffiliation{on leave from Ioffe Physico-Technical Institute, St. Petersburg}
\affiliation{nMat Group, CNRS/CEMES, 29 Rue Jeanne Marvig, 31055 Toulouse, France}
\author{G. Ben Assayag}
\affiliation{nMat Group, CNRS/CEMES, 29 Rue Jeanne Marvig, 31055 Toulouse, France}
\author{S. Schamm}
\affiliation{nMat Group, CNRS/CEMES, 29 Rue Jeanne Marvig, 31055 Toulouse, France}
\author{G. Zanchi}
\affiliation{nMat Group, CNRS/CEMES, 29 Rue Jeanne Marvig, 31055 Toulouse, France}
\author{C. Colliex}
\affiliation{Laboratoire de Physique des Solides, Universit\'{e} Paris-Sud - UMR 8502, 91405 Orsay, France}
\author{W. M\"{o}ller}
\affiliation{Research Center Rossendorf, Institute of Ion Beam Physics and Materials Research, PO
Box 51 01 19, 01314 Dresden, Germany}
\author{A. Claverie}
\affiliation{nMat Group, CNRS/CEMES, 29 Rue Jeanne Marvig, 31055 Toulouse, France}
\keywords{Nonvolatile Memory, Si nanocrystals, PEELS-STEM, Plasmon mapping}

\begin{abstract}
Scalability and performance of current flash memories can be improved substantially by replacing the floating poly-Si
gate by a layer of Si dots. This multi-dot layer can be fabricated CMOS-compatibly in very thin gate oxide by ion beam
synthesis (IBS). Here, we present both experimental and theoretical studies on IBS of multi-dot layers consisting of Si
nanocrystals (NCs). The NCs are produced by ultra low energy Si$^{+}$ ion implantation, which causes a high Si
supersaturation in the shallow implantation region. During post-implantation annealing, this supersaturation leads to
phase separation of the excess Si from the SiO$_{2}$. Till now, the study of this phase separation process suffered
from the weak Z contrast between Si and SiO$_{2}$ in Transmission Electron Microscopy (TEM). Here, this imaging problem
is resolved by mapping Si plasmon losses with a Scanning Transmission Electron Microscopy equipped with a parallel
Electron Energy Loss Spectroscopy system (PEELS-STEM). Additionally, kinetic lattice Monte Carlo simulations of Si
phase separation have been performed and compared with the experimental Si plasmon maps. It has been predicted
theoretically that the morphology of the multi-dot Si floating-gate changes with increasing ion fluence from isolated,
spherical NCs to percolated spinodal Si pattern. These patterns agree remarkably with PEELS-STEM images. However, the
predicted fluence for spinodal patterns is lower than the experimental one. Because oxidants of the ambient atmosphere
penetrate into the as-implanted SiO$_{2}$, a substantial fraction of the implanted Si might be lost due to oxidation.
\end{abstract}
\date{accepted APL draft, 4$^{th}$ Version, \today}
\maketitle

Metal-Oxide-Silicon Field-Effect-Transistors (MOSFETs) with an electrically isolated (\textquotedblleft
floating\textquotedblright) gate layer embedded in the gate oxide are currently used as flash memories. The replacement
of this floating-gate by a layer of discrete Si nanocrystals (NCs) \cite{Tiwari1996} improves the performance of flash
memories substantially \cite{Tiwari2000}. The reduced probability for a complete discharging of the multi-dot
floating-gate by oxide defects allows thinner tunnel oxides. In turn, the floating-gate will be charged/discharged by
quantum mechanical direct electron tunneling (instead of defect-generating Fowler-Nordheim tunneling). The memory
operation voltage can be reduced and scalability is improved. Using ion beam synthesis, the multi-dot floating-gate can
be fabricated along with standard CMOS processing \cite{Kapetanakis2002}. Si$^{+}$ ions are implanted at ultra low
energies into the gate oxide, causing there a high Si supersaturation. During post-implantation annealing, this Si
supersaturation leads to phase separation of elemental Si from SiO$_{2}$ \cite{Mueller2002}. Imaging this phase
separation process is difficult. Till now, Transmission Electron Microscopy (TEM) has suffered from weak Z contrast
between Si and SiO$_{2}$ phases. Recently, this problem was partially overcome by Fresnel imaging using under-focused
bright field conditions \cite{Assayag2003}. \enlargethispage*{3em} Thus, the distance of the layer of phase separated
Si from the transistor channel could be determined \cite{Carrada2003,Bonafos2004}. However, this technique fails to
resolve the morphology of the phase separated Si. For instance, recent kinetic Monte Carlo (KMC) simulations of phase
separation predict a pronounced fluence dependence of the precipitate morphology \cite{Mueller2002}. For low Si$^{+}$
fluences, spherical and isolated Si NCs form by nucleation and growth, while for higher Si$^{+}$ fluences spinodal
decomposition occurs. The elongated, non-spherical Si structures, formed by spinodal decomposition, coalesce at even
higher fluences to an interconnected, labyrinthine Si network. Here, using a Scanning TEM (STEM) with an efficient
parallel Electron Energy Loss Spectroscopy (PEELS) system, the predictions of KMC simulations are confirmed for the
first time. The contrast problem of conventional TEM could be overcome by mapping Si plasmon losses, which differ from
SiO$_2$. Comprehensive KMC studies and PEELS-STEM analysis has been performed to understand the complex process of
phase separation in a thin buried layer.

To form Si NCs by phase separation, Si$^{+}$ ions have been implanted at $1\operatorname*{keV}$ energy into
$10\operatorname{nm}$ thick SiO$_{2}$ layers, which were thermally grown on (001) Si substrates. Using an AXCELIS
GSD-ULTRA ultra-low-energy implanter, fluences of $5\times10^{15}\operatorname{cm}^{-2}$,
$1\times10^{16}\operatorname{cm}^{-2}$, and $2\times10^{16}\operatorname{cm}^{-2}$ were implanted at room temperature.
Surface charging due to implantation was compensated by a Xe plasma electron flood gun. The implanted samples were
cleaned using a piranha solution and furnace annealed for $30\operatorname{min}$ in N$_{2}$ at
$950\operatorname{{}^{\circ}{\rm C}}$.
\begin{figure}[t]
\begin{center}
\includegraphics[width=86mm]{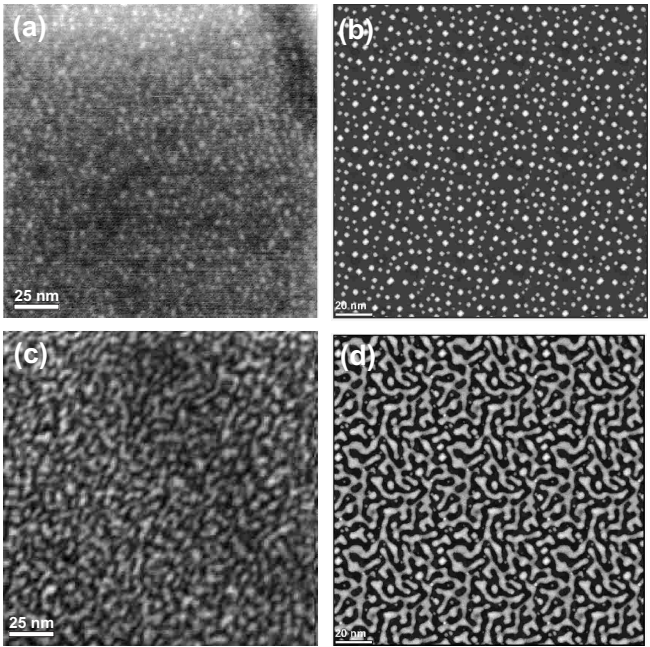}
\caption{Morphology of phase separated Si in SiO$_{2}$. Si plasmon maps by PEELS-STEM (a,c) are compared to
corresponding KMC simulations (b,d). Nucleation of Si NCs is observed (a) for a Si fluence of
$1\times10^{16}\operatorname{cm}^{-2}$ and predicted with the same morphology for (b)
$3\times10^{15}\operatorname{cm}^{-2}$. Spinodal patterns are imaged (c) for $2\times10^{16}\operatorname{cm}^{-2}$ and
simulated (d) for $8\times10^{15}\operatorname{cm}^{-2}$. White and black regions correspond to Si and SiO$_{2}$
phases, respectively.} \label{Plane View}
\end{center}
\end{figure}
From all samples, cross sectional and plane view TEM specimens were prepared by the standard procedure of grinding,
dimpling and Ar$^{+}$ ion beam thinning. PEELS-STEM was performed on plane view samples using a Scanning TEM VG-HB 501
operating at $100\operatorname{kV}$ that is equipped with a field emission cathode and a parallel Gatan 666 EELS
spectrometer. Low-loss EELS spectra were recorded at each picture point, hence a spectrum image was acquired
\cite{Jeangillaume1989}. Using a non negative least square method \cite{Lawson1974}, these spectra were fitted in the
energy range of $15-30\operatorname{eV}$ by a weighted sum of two plasmon reference spectra, which were obtained from
bulk Si and SiO$_{2}$. Potential size-effects like shifts of the Si plasmon resonance of small Si NCs were not
compensated. The gray level of the PEELS-STEM image pixels is given by the ratio of the intensity of the Si plasmon
peak and the total intensity of the Si and SiO$_{2}$ plasmon peaks and, therefore, maps the Si bulk plasmons.
Quantitative Si concentrations, however, can not be given, only the relative Si content is imaged. In Fig.~\ref{Plane
View}, such Si plasmon maps (on the left hand side) are compared to plane view snapshots of kinetic Monte Carlo
simulations (on the right hand side). This atomistic approach to phase separation of excess Si in thin SiO$_{2}$ layers
by atomistic simulations was recently described in detail in Ref.~\cite{Mueller2002} and \cite{Mueller2003}. The depth
profiles of excess Si  due to high fluence Si$^{+}$ ion implantation into thin gate oxides have been calculated by the
binary collision program TRIDYN \cite{Moeller1984} accounting for the effects of ion erosion, target swelling and ion
beam mixing dynamically. The post-implantation phase separation during thermal treatment is described by a kinetic 3D
lattice Monte Carlo program package \cite{Heinig2003,Strobel2001}. \enlargethispage*{3em} Taking the TRIDYN profiles of
Si excess as well as Si solubility and diffusivity in SiO$_{2}$ as input, the program describes excess Si diffusion,
precipitation and Ostwald ripening in the thin SiO$_{2}$ layer under the constraints of the boundary conditions of a
nearby Si/SiO$_{2}$ interface and a free SiO$_{2}$ surface. Here, we use a simplified version of the KMC program,
i.e.~only Ising-type nearest-neighbor interactions of diffusing Si atoms. It should be noted that KMC simulations with
measured Si self-diffusivities \cite{Uematsu2004,Mathiot2003} lead to too long annealing times or too high
temperatures. Obviously, the diffusive Si mass transport by a mobile SiO$_2$ defect with local Si excess (that could
either be named Si interstitial, SiO molecule \cite{Uematsu2004}, or oxygen vacancy \cite{Song2001}) does not
necessarily follow the same mechanism than as the $^{28}$SiO$_2$/$^{30}$SiO$_2$ interface broadening, which was
analyzed in self-diffusivity studies. In this letter, the discrepancy between diffusive Si mass transport and Si
self-diffusivity will not be discussed as our theoretical predictions aim at the reaction pathway of SiO$_x$
decomposition rather than at a quantitative prediction of annealing time and temperature. Thus, KMC simulations utilize
a relative time scale "Monte Carlo steps" (MCS) that allows a posterior recalibration with realistic Si diffusivities
\cite{Strobel2001}.

Subsequent image processing of KMC simulation data allowed to obtain KMC simulation snapshots (plane view or cross
section) that can be compared directly to PEELS-STEM or Fresnel images. In plane view, the simulation cell has been
tripled laterally taking advantage of the periodic boundary conditions. The number of excess Si atoms in the vertical
column below the pixel at $(x,y)$ determines the gray level of that pixel. The highest occurring number of Si atoms in
such a column is assigned to white, whereas black means that no Si excess is found in the SiO$_{2}$ matrix. In plane
view images, Si atoms of the underlying (001) Si substrate are not considered.
\begin{figure}[t]
\begin{center}
\includegraphics[width=86mm]{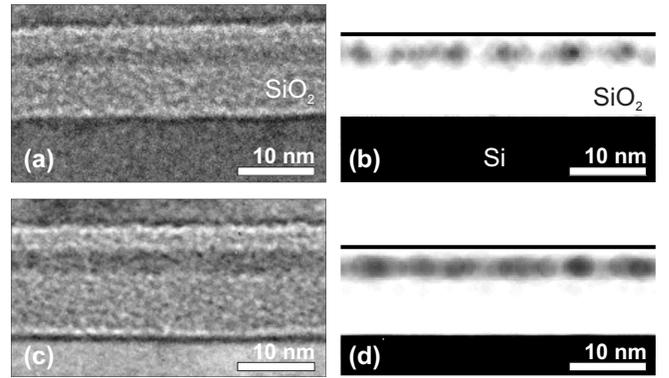}
\caption{Cross section view of the layer of phase separated Si in SiO$_{2}$. Fresnel XTEM images
for (a) $1\times10^{16}$Si$^{+}\operatorname{cm}^{-2}$ and (c)
$2\times10^{16}$Si$^{+}\operatorname{cm}^{-2}$ are compared to cross sectional KMC simulation
snapshots for (b) $3\times10^{15}$Si$^{+}\operatorname{cm}^{-2}$ and (d)
$8\times10^{15}$Si$^{+}\operatorname{cm}^{-2}$.}
\label{Cross section}%
\end{center}
\end{figure}

No Si precipitates were observed in the sample for the lowest Si$^{+}$ fluence of
$5\times10^{15}\operatorname{cm}^{-2}$ (TEM images not shown here). For the medium Si$^{+}$ fluence of
$1\times10^{16}\operatorname{cm}^{-2}$, white spots in a dark background appear in the plane view Si plasmon map shown
in Fig.~\ref{Plane View} (a) indicating spherical Si NCs embedded in the SiO$_{2}$ matrix. During thermal annealing,
these NCs have formed by nucleation and growth in the Si enriched region of the implanted SiO$_{2}$ layer. The observed
NCs are relatively small with an estimated mean diameter of 2.7 nm and have a high area density of
$3.3\times10^{12}\operatorname{cm}^{-2}$. A similar precipitate morphology with the same mean NC size and density is
found in the plane view KMC simulation snapshot of Fig.~\ref{Plane View} (b), but for a significantly lower Si$^{+}$
fluence of $3\times10^{15}\operatorname{cm}^{-2}$ and a KMC simulation (annealing) time of $2100\operatorname*{kMCS}$.
It should be noted that even this \textquotedblleft medium\textquotedblright\ experimental Si$^{+}$ fluence of
$1\times10^{16}\operatorname{cm}^{-2}$ leads in the KMC simulation to the formation of coalesced poly-Si layer buried
in the SiO$_{2}$ \cite{Mueller2002}. For the highest Si$^{+}$ fluence of
$2\times10^{16}$Si$^{+}\operatorname{cm}^{-2}$, the morphology of the phases separated Si has changed completely as
observed in Fig.~\ref{Plane View} (c). A spaghetti-like network of white and black regions is found. This spinodal
pattern clearly indicate that, at this fluence, phase separation took place by spinodal decomposition. Even more,
percolation is observed. Si precipitates are no longer spatially isolated, but an interconnected network of phase
separated Si has formed. An equivalent morphology with the same typical distances than in Fig.~\ref{Plane View} (c)
between the spinodal fingers is found in the plane view KMC simulation snapshot of Fig.~\ref{Plane View} (d). This
results was obtained for $8\times10^{15}$ Si$^{+}\operatorname{cm}^{-2}$ and a simulation time of
$300\operatorname*{kMCS}$. Strikingly, the morphology agrees remarkably well between both images, while simulation
snapshots for other fluences or annealing times deviate instead considerably from the morphology seen in
Fig.~\ref{Plane View} (c).

Samples were studied additionally in cross section using a CM30 Phillips TEM equipped with a LaB$_6$ cathode operating
at $300\operatorname{kV}$. Fresnel imaging conditions were applied in order to achieve at least a weak contrast between
Si and SiO$_{2}$ in cross sectional TEM (XTEM) images \cite{Assayag2003}. I.e. images were taken at Out-of-Bragg
alignment and under strongly underfocused bright field conditions. Details of the responsible contrast mechanism are
given in Ref.~\cite{Assayag2003}. In Fig.~\ref{Cross section}, (a) and (c) show the XTEM Fresnel images for fluences of
$1\times10^{16}$ Si$^{+}\operatorname{cm}^{-2}$ and $2\times10^{16}$ Si$^{+}\operatorname{cm}^{-2}$, respectively.
Cross section KMC simulation snapshots are displayed in Fig.~\ref{Cross section} (b) and (d) for $3\times10^{15}$
Si$^{+}\operatorname{cm}^{-2}$ ($2100\operatorname*{kMCS}$) and $8\times10^{15}$ Si$^{+}\operatorname{cm}^{-2}$
($300\operatorname*{kMCS}$), respectively. Thereby, the same type of image processing was applied to KMC cross section
views than to plane views. However, substrate atoms are considered here too and the gray scale of the KMC images has
been inverted. White and black regions correspond to SiO$_{2}$ and Si phases, respectively. In all images, the phase
separated Si forms a single, sharp layer seen as gray band in the SiO$_{2}$ that is well separated from the
SiO$_{2}$/Si interface. Due to the extremely shallow Si excess profile obtained by low-energy Si$^{+}$ implantation,
phase separation is quasi confined to two dimensions. For $1\times10^{16}$ Si$^{+}\operatorname{cm}^{-2}$, Si
precipitates align nicely in a thin layer, which is just a few nanometers thick, Fig.~\ref{Cross section} (a). When the
Si$^{+}$ fluences is increased to $2\times10^{16}$ Si$^{+}\operatorname{cm}^{-2}$, the precipitate layer remains
comparatively well localized in depth, Fig.~\ref{Cross section} (c), although the Si morphology observed in plane view
has changed completely,  Fig.~\ref{Plane View} (c). The total SiO$_{2}$ thickness is systematically smaller for the KMC
simulations than for the corresponding Fresnel XTEM images. Nevertheless, the distance between the Si/SiO$_{2}$
interface and the phase separated Si agrees nicely for experiment and simulation.

\enlargethispage*{3em} The Si fluences in our KMC simulations where chosen in order to obtain morphologies of phase
separated Si, which are similar to experimental ones. This adjustments of our simulation reveals a strong discrepancy
between experimental and theoretical Si fluences. More Si than theoretically predicted has to be implanted. Not all Si
that has nominally been implanted into the SiO$_{2}$ is available for phase separation. The reason for the missing Si
excess might be twofold. (i) Recent Time of Flight Secondary Ion Mass Spectroscopy measurements on low-energy,
low-fluence $^{30}$Si$^{+}$ as-implanted SiO$_{2}$ samples indicate that only a fraction of about $0.5-0.7$ of the
nominal Si$^{+}$ fluence has been implanted into the SiO$_{2}$ \cite{Perego2003}.
(ii) It is known that (Si or Ge) NC formation in very thin SiO$_{2}$ films is extremely sensitive to humidity absorbed
by the as-implanted (damaged) glass network \cite{Schmidt2002} as well as to oxidants being present either in the
annealing ambient \cite{Oswald2000}. Thus, a considerable amount of the implanted Si might become oxidized during
annealing. This explains also why no NCs have been observed by TEM for the lowest Si$^{+}$ fluence of
$5\times10^{15}\operatorname{cm}^{-2}$. The implanted Si has been oxidized completely during the thermal annealing. At
the same time, oxidation of the implanted Si leads to a volume expansion that increases the overall SiO$_{2}$ layer
thickness \cite{Carrada2003}. This swelling of the SiO$_{2}$ due to Si oxidation can be seen in the XTEM images of
Fig.~\ref{Cross section} (a,c) if compared to the KMC simulation snapshots Fig.~\ref{Cross section} (b,d), which just
include the SiO$_{2}$ expansion due to the incorporated Si atoms \cite{Mueller2003}. Here, the KMC simulations do not
account for oxidation and the swelling caused by it. To do so, a multi-component KMC approach (Si + O) is needed.
Though samples have been annealed under fixed experimental conditions, the KMC simulation snapshots of corresponding Si
patterns refer to different simulation times. Two reasons might be responsible for this discrepancy in the evolution
speed. At first, the oxidation of a substantial part of implanted Si might influence the kinetics, and secondly, the Si
bulk diffusion in SiO$_{2}$ might differ substantially from the Si diffusion at the Si/SiO$_{2}$ interface, which is
assumed till now in the simulations. Studies that investigate this point are underway.

Summarizing, extensive studies on low-energy ion beam synthesis of multi-dot Si floating-gates embedded in thin
SiO$_{2}$ layers have been presented. The morphology of this floating-gate layer formed by Si phase separation from
SiO$_{2}$ has been revealed by Si plasmon mapping using PEELS-STEM. A direct comparison to kinetic 3D lattice Monte
Carlo simulation snapshots have been made for the first time and shows a remarkable agreement between the atomistic
simulations and the PEELS-STEM images. A strong fluence dependence of the precipitate morphology is confirmed. For low
Si$^{+}$ fluences, isolated Si NCs form by nucleation and growth, while high fluences lead to spinodal decomposition
during annealing and therefore to the formation of elongated precipitates, which additionally become percolated. With
respect to a future applications in nonvolatile multi-dot floating-gate memories, structural percolation of the Si
precipitates should be avoided. Otherwise, electrical charge brought to interconnected Si precipitates could spread
easily in lateral dimension, i.e. the layer of phase separated Si would behave like a conventional poly-Si
floating-gate. The present studies point out that, in order to prevent percolation, Si implantation (at
$1\operatorname*{keV}$ energy) should not exceed a limiting fluence of $1\times10^{16}$ Si$^{+}$cm$^{-2}$. Then, a thin
layer of isolated, spherical Si NCs forms during annealing. However, the predicted fluence for spinodal pattern is
lower than the experimental one. A substantial fraction of the implanted Si might be lost due to oxidation by oxidants
penetrated from the ambient atmosphere.

This work was supported by the European Commission through the Growth project G5RD/2000/00320 --
NEON (Nanoparticles for Electronics).


\begin{thebibliography}{99}
\bibitem {Tiwari1996}S. Tiwari, F. Rana, H. Hanafi, A. Hartstein, E. F. Crabbe,
                     and K. Chan, Appl. Phys. Lett. 68, 1377 (1996).
\bibitem {Tiwari2000}S. Tiwari, J.A. Wahl, H. Silva, F. Rana, J.J. Welser, Appl. Phys. A 71 403 (2000).
\bibitem {Kapetanakis2002}E. Kapetanakis, P. Normand, D. Tsoukalas, K. Beltsios, Appl. Phys. Lett. 80, 2794 (2002).
\bibitem {Mueller2002}T. M\"{u}ller, K.-H. Heinig, and W. M\"{o}ller, Appl. Phys. Lett. 81, 3049 (2002).
\bibitem {Assayag2003}G. B. Assayag, C. Bonafos, M. Carrada, P. Normand, D. Tsoukalas,
                        and A. Claverie, Appl. Phys. Lett. 82, 200 (2003).
\bibitem {Carrada2003}M. Carrada, N. Cherkashin, C. Bonafos, G. Benassayag,
                        D. Chassaing, P. Normand, D. Tsoukalas, V. Soncini, A. Claverie, Mat. Sci. \& Eng. B 101 204 (2003).
\bibitem {Bonafos2004}C. Bonafos, M. Carrada, N. Cherkashin, H. Coffin, D. Chassaing, G. Ben Assayag, A. Claverie,
                     T. M\"{u}ller, K. H. Heinig, M. Perego, M. Fanciulli, P. Normand, and D. Tsoukalas, J. Appl.
                     Phys. 95, 5696 (2004).
\bibitem {Jeangillaume1989}C. Jeanguillaume and C. Colliex, Ultramicroscopy 28, 252 (1989).
\bibitem {Lawson1974}C.L. Lawson, R.J. Hanson, Solving least square problems, Prentice-Hall, Englewood cliffs,
                     New-Jersey, 1974.
\bibitem {Mueller2003}T. M\"{u}ller, K.-H. Heinig, and W. M\"{o}ller, Mat. Sci. \& Eng. B 101/1-3, 49 (2003).
\bibitem {Moeller1984}W. M\"{o}ller, W. Eckstein, Nucl Instr. \& Meth. in Phys. Res. B 2 814 (1984).
\bibitem {Heinig2003}K.-H. Heinig, T. M\"{u}ller, B. Schmidt, M. Strobel, W. M\"{o}ller, Appl. Phys. A, 77 (2003) 17.
\bibitem {Strobel2001}M. Strobel, K.-H. Heinig, and W. M\"{o}ller, Phys. Rev. B 64, 245422 (2001).
\bibitem {Uematsu2004} M. Uematsu, H. Kageshima, Y. Takahashi, S. Fukatsu, K. M. Itoh, K. Shiraishi, and U. G\"{o}sele,
                     Appl. Phys. Lett. 84, 876 (2004).
\bibitem {Mathiot2003} D. Mathiot, J. P. Schunck, M. Perego, M. Fanciulli, P. Normand, C. Tsamis, and D. Tsoukalas,
                     J. Appl. Phys. 94, 2136 (2003).
\bibitem {Song2001}J. Song, L. R. Corrales, G. Kresse, and H. Jonsson, Phys. Rev. B 64, 134102 (2001).
\bibitem {Perego2003}M. Perego, M. Fanculli, G. Ben Assayag, A. Claverie, private communication (2003).
\bibitem {Oswald2000}S. Oswald, B. Schmidt, K.-H. Heinig, Surf. Interface Anal. 29, 249 (2000).
\bibitem {Schmidt2002}B. Schmidt, D. Grambole, F. Herrmann, Nucl Instr. \& Meth. in Phys. Res. B 191 482 (2002).
\end{thebibliography}
\end{document}